\documentclass{ws-procs10x7-notrim}
\usepackage{balance}

\def\bfsigma{\mbox{\boldmath $\sigma$}}

\def\lQ{\Lambda_{\rm QCD}}

\newcommand{\be}{\begin{equation}}
\newcommand{\ee}{\end{equation}}
\newcommand{\bea}{\begin{eqnarray}}
\newcommand{\eea}{\end{eqnarray}}

\def\als{\alpha_{\rm s}}
\def\siml{{\ \lower-1.2pt\vbox{\hbox{\rlap{$<$}\lower6pt\vbox{\hbox{$\sim$}}}}\ }} 
\def\simg{{\ \lower-1.2pt\vbox{\hbox{\rlap{$>$}\lower6pt\vbox{\hbox{$\sim$}}}}\ }}

\def\cite#1{[\refcite{#1}]}

\makeindex
\begin{document}

\title{HEAVY HADRON SPECTROSCOPY}

\author{ANTONIO VAIRO}

\address{Dipartimento di Fisica dell'Universit\`a di Milano and INFN, \\ 
via Celoria 16, 20133 Milano, Italy \\
E-mail: antonio.vairo@mi.infn.it}

\twocolumn[\maketitle\abstract{
I review recent theoretical advances in heavy hadron spectroscopy.
}
]

\section{Introduction}
After having played a major role in the foundation of QCD, heavy hadron spectroscopy 
has witnessed in the last years a renewal of interest led by the many new data 
coming from the B factories, CLEO and the Tevatron and by the progress 
made in the theoretical methods. I will summarize the former and mostly focus on the latter. 
Much of the theoretical progress in the physics of heavy hadrons comes from effective field theories (EFTs) 
and lattice gauge theories. For most of these systems, they allow systematic treatments,  
which may be used to gain control over one of the most elusive sectors 
of the Standard Model, low-energy QCD, even in one of its most spectacular manifestations:
the formation of exotic bound states.

A systematic treatment of heavy-quark bound states is possible because the systems 
are characterized by at least two small parameters. One is the strong coupling 
constant at the heavy-quark mass scale, $m$, which is, by definition, larger than the typical 
hadronic scale, $\lQ$, and the other is the ratio $\lQ/m$.
Expansions in these two parameters can be exploited in the description of systems made by one 
heavy quark, like heavy-light mesons or baryons. If the expansions are made manifest at the Lagrangian 
level, the resulting EFT is known as Heavy Quark Effective Theory, HQET \cite{Isgur:1989vq}  
(for some reviews see \cite{Neubert:1993mb}).

Systems made by two heavy quarks are most frequently and successfully studied as 
non-relativistic bound states \cite{Brambilla:2004wf}.
They are characterized by another small parameter, the heavy-quark velocity $v$,
which comes with a hierarchy of energy scales: $m v$, $m v^2$, ...
Making explicit at the Lagrangian level the expansions in $mv/m$ and $mv^2/m$ leads 
to an EFT known as non-relativistic QCD (NRQCD) \cite{Caswell:1985ui,Bodwin:1994jh}.
This EFT is similar to HQET, but with a different power counting.
It also accounts for contact interactions between quarks and antiquarks (e.g. in decay processes) and hence has a wider 
set of operators. Making explicit at the Lagrangian level the expansion in $mv^2/mv$ 
leads to another EFT known as potential NRQCD (pNRQCD) \cite{Pineda:1997bj}
(an alternative EFT is in \cite{Luke:1999kz}).
pNRQCD is close to a Schr\"odinger-like description of the bound
state and hence as simple. The bulk of the interaction
is carried by potential-like terms, but non-potential interactions,
associated with the propagation of low-energy degrees of freedom, are generally present as well.
For a review on non-relativistic EFTs we refer to \cite{Brambilla:2004jw}. 

It is important to establish when $\lQ$ sets in, i.e. when we have to 
resort to non-perturbative methods. In the case of systems made 
by one heavy quark, $\lQ$ becomes the most relevant scale once the heavy-quark 
mass has been integrated out: all HQET matrix elements are non-perturbative.
The situation is more variegated in systems made by two heavy quarks.
For low-lying resonances, it is reasonable, although not proved, to assume 
$mv^2 \simg \lQ$. The system is weakly coupled and we may rely on perturbation theory,
for instance, to calculate the potential. The theoretical challenge here is performing higher-order 
calculations and the goal is precision physics. For high-lying resonances, we assume
$mv \sim \lQ$.  The system is strongly coupled and the potential must be determined 
non-perturbatively, for instance, on the lattice.  The theoretical challenge here is providing a consistent framework 
where to perform lattice calculations and the progress is measured by the advance in lattice computations.

For what concerns systems close or above the open flavor threshold, 
a complete and satisfactory understanding of the dynamics has not been achieved so far. 
Hence, the study of these systems is on a less secure ground than the study of states below threshold. 
Although in some cases one may develop an EFT owing to special dynamical conditions
(an example that we will discuss in the following is the $X(3872)$ interpreted as a 
loosely bound $D^0 \, \bar{D}^{*\,0}$ $+$ ${\bar D}^0 \, D^{*\,0}$ molecule), 
the study of these systems largely relies on phenomenological models. 
The major theoretical challenge here is to interpret the new 
states in the charmonium region discovered at the B-factories in the last years 
(for an updated list, see the Quarkonium Working Group (QWG) page {\sf http://www.qwg.to.infn.it/}).

In the following, I will discuss from a theoretical perspective advances in heavy hadron 
spectroscopy in the light of the results presented at ICHEP 2006.
For charm and charmonium in experiments I refer to \cite{Pakhlov:ICHEP06,Li:ICHEP06}, for light quark spectroscopy 
to \cite{Zaitsev:ICHEP06}. An overview on the newly discovered charmonium resonances from an experimental 
perspective can be found in \cite{Mussa:ICHEP06}.

\section{Heavy-quark masses}
The heavy-quark masses are among the fundamental parameters of the Standard Model that may be 
extracted with precision from heavy-light or heavy-heavy quark systems.
Different recent determinations of the $c$ mass 
(from the $J/\psi$, ${\overline m}_c({\overline m}_c) = 1.24 \pm 0.02$ GeV \cite{Brambilla:2001fw}, 
from the $D$, ${\overline m}_c({\overline m}_c) = 1.21\pm 0.1$ GeV \cite{Pineda:2001zq}, 
from high moments sum rules, ${\overline m}_c({\overline m}_c) = 1.19\pm 0.11$ GeV \cite{Eidemuller:2002wk}, 
from inclusive $B$ decays,  ${\overline m}_c({\overline m}_c) = 1.224\pm 0.06$ GeV \cite{Hoang:2005zw}, 
from low moments sum rules, ${\overline m}_c({\overline m}_c) = 1.290\pm 0.015$ GeV \cite{Kuhn:ICHEP06})
and the $b$ mass (from the $\Upsilon(1S)$, ${\overline m}_b({\overline m}_b) = 4.19 \pm 0.03$ GeV \cite{Brambilla:2001qk},
${\overline m}_b({\overline m}_b) = 4.346 \pm 0.070$ GeV \cite{Penin:2002zv}, 
${\overline m}_b({\overline m}_b) = 4.20 \pm 0.04$ GeV \cite{Lee:2003hh}, 
${\overline m}_b({\overline m}_b) = 4.24 \pm 0.07$ GeV \cite{Contreras:2003zb},
from  $B\to X_s\gamma$, ${\overline m}_b({\overline m}_b) = 4.22 \pm 0.09$ GeV \cite{Grinstein:2001yg},
from high moments sum rules, ${\overline m}_b({\overline m}_b) = 4.19 \pm 0.06$ GeV \cite{Pineda:2006gx},
from low moments sum rules, ${\overline m}_b({\overline m}_b) = 4.18 \pm 0.035$ GeV \cite{Kuhn:ICHEP06})
show a remarkable agreement (with the possible exception of \cite{Penin:2002zv}, for a critical 
discussion we refer to \cite{Brambilla:2004wf}).

\section{Heavy-light mesons and baryons}
$\bar{Q}q$ and $Qqq$ systems are characterized by the hierarchy of scales $m \gg \lQ$, 
where the typical distance of the heavy quark from the light ones is of order $1/\lQ$. 
Theoretical advances in the calculation of heavy hadron lifetimes 
in the framework of HQET have been presented in \cite{Petrov:ICHEP06}. 

In the last year, new $D_{sJ}$ searches have led to the discovery of new
resonances at BELLE, $D_{sJ}(2700)$, $\Gamma \approx 115~{\rm MeV}$
\cite{Brodzicka:ICHEP06}, and at BABAR, $D_{sJ}(2856)$, $\Gamma \approx 47~{\rm MeV}$ \cite{delRe:ICHEP06}, and 
new $B_{sJ}$  searches to the discovery of a new resonance at CDF,  
$B_{s1}(5829)$ \cite{Gorelov:ICHEP06}. $cqq$ baryons  have been investigated at BABAR
leading to the discovery of the $\Lambda_c(2940)$, $\Gamma \! \approx \!
17$ MeV \cite{Petersen:ICHEP06,Kim:ICHEP06} and at BELLE leading to the discovery of the 
 $\Xi_c(2980)^+$, $\Gamma \approx 43.5~{\rm MeV}$ and the $\Sigma_c(3077)^+$, $\Gamma \approx 5.2~{\rm MeV}$
\cite{Mizuk:ICHEP06}.

\section{Low-lying $Q\bar{Q}$}
Low-lying $Q\bar{Q}$ states are expected to realize the hierarchy: $ m \gg  mv
\gg mv^2 \simg \lQ$, where $mv$ is the typical scale of the inverse distance between the
heavy quark and antiquark and $mv^2$ the typical scale of the binding energy.
At a scale $\mu$ such that $mv \gg \mu \gg mv^2$ the effective degrees of freedom are 
$Q\bar Q$ states (in color singlet and octet configurations), low-energy gluons and light quarks. 

The lowest-lying quarkonium states are $\eta_b$ (not yet detected), $\Upsilon(1S)$, $\eta_c$,  
$J/\psi$, $B_c$ and $B_c^*$ (not yet detected). 
As mentioned above, the $\Upsilon(1S)$ and $J/\psi$ masses may be used to extract  
the bottom and charm quark masses. Once the heavy-quark masses are known, one
may use them for the determination of quarkonium ground-state observables. At NNLO the $B_c$ mass was calculated 
in \cite{Brambilla:2000db} ($M_{B_c} = 6326 \pm 29$  MeV), 
\cite{Brambilla:2001fw} ($M_{B_c} = 6324 \pm 22$  MeV) and 
\cite{Brambilla:2001qk} ($M_{B_c} = 6307 \pm 17$  MeV). These values 
agree well with the unquenched lattice determination of \cite{Allison:2004be} 
($M_{B_c} = 6304 \pm 12^{+18}_{-0}$ MeV), which shows that the $B_c$ mass
is not very sensitive to non-perturbative effects. 
This is confirmed by a recent measurement of the $B_c$ in the channel  $B_c \to J/\psi \, \pi$ 
by the CDF collaboration at the Tevatron; they obtain with 360 pb$^{-1}$ of data 
$M_{B_c} = 6285.7 \pm 5.3 \pm 1.2$ MeV \cite{Acosta:2005us}, while the latest
available figure based on 1.1 fb$^{-1}$ of data is 
$M_{B_c} = 6276.5 \pm 4.0 \pm 2.7$ MeV 
(see {\sf $\hbox{\sf http:}$//www-cdf.fnal.gov/physics/new/bottom/060525.ble\-ssed-bc-mass/}).

The bottomonium (and charmonium) ground-state hyperfine splitting has been calculated at NLL
in \cite{Kniehl:2003ap}. Combining it with the measured $\Upsilon(1S)$ mass,
this determination provides a quite precise prediction for the $\eta_b$ mass: 
$M_{\eta_b} = 9421 \pm 10^{+9}_{-8} ~{\rm MeV}$, where the first error is an estimate of the
theoretical uncertainty and the second one reflects the uncertainty in $\als$.
Note that the discovery of the $\eta_b$ may provide a very competitive 
source of $\als$ at the bottom mass scale with a projected error at the $M_Z$ 
scale of about $0.003$. Similarly, in \cite{Penin:2004xi}, 
the hyperfine splitting of the $B_c$ was calculated at NLL
accuracy: $M_{B_c^*}  - M_{B_c} = 65 \pm 24^{+19}_{-16}~{\rm MeV}$.

The ratios of electromagnetic decay widths were calculated for the ground state 
of charmonium and bottomonium at NNLL order in \cite{Penin:2004ay}; 
for the latter the result is $\Gamma(\eta_b\to\gamma\gamma) /\Gamma(\Upsilon(1S)\to e^+e^-) = 0.502
\pm 0.068 \pm 0.014$. A partial  NNLL order analysis for the absolute width of $\Upsilon(1S) \to
e^+e^-$ can be found in \cite{Pineda:2006ri}.

Allowed magnetic dipole transitions between charmonium and bottomonium ground states
have been considered at NNLO in \cite{Brambilla:2005zw,Vairo:2006js}.
The results are: $\Gamma(J/\psi \to \gamma \, \eta_c) \! = (1.5 \pm 1.0)~\hbox{keV}$
and $\Gamma(\Upsilon(1S) \to \gamma\,\eta_b)$ $=$  $(k_\gamma/39$ $\hbox{MeV})^3$
$\,(2.50 \pm 0.25)$ $\hbox{eV}$, where the errors account for uncertainties coming from higher-order corrections.
The width $\Gamma(J/\psi \to \gamma\,\eta_c)$ is consistent with the PDG value 
\cite{Yao:2006px}. Concerning $\Gamma(\Upsilon(1S) \to \gamma\,\eta_b)$, 
a photon energy $k_\gamma = 39$ MeV corresponds to a $\eta_b$ mass of 9421 MeV. 

The radiative transition $\Upsilon(1S)\to\gamma\,X$ has been considered in 
\cite{Fleming:2002sr,GarciaiTormo:2005ch}. The agreement with the CLEO data  
\cite{Nemati:1996xy} is very good.

\section{Low-lying $QQq$}
The SELEX collaboration at Fermilab reported evidence of five resonances that 
may be possibly identified with doubly charmed baryons \cite{Ocherashvili:2004hi}. 
Although these findings have not been confirmed by other experiments (notably 
by BELLE \cite{Lesiak:2006sk} and BABAR \cite{Aubert:2006qw,Kim:ICHEP06}) 
they have triggered a renewed theoretical interest in doubly heavy baryon systems.

Low-lying $QQq$ states are expected to realize the hierarchy: $ m \gg mv \gg
\lQ$, where $mv$ is the typical inverse distance between the two heavy quarks
and $\lQ$ is the typical inverse distance between the center-of-mass of the two heavy quarks
and the light quark. 

At a scale $\mu$ such that $mv \gg \mu \gg \lQ$ the effective  
degrees of freedom are $QQ$ states (in color antitriplet and sextet configurations), 
low-energy gluons and light quarks. Since the system shares features 
of heavy-light mesons and quarkonia, the most suitable EFT at
the scale $\mu$ is a combination of pNRQCD and HQET \cite{Brambilla:2005yk,Fleming:2005pd}. 
The hyperfine splittings of the doubly heavy 
baryon lowest states have been calculated at NLO in $\als$ and at LO in
$\lQ/m$ by relating them to the hyperfine splittings of the $D$ and $B$ mesons (this 
method was first used in \cite{Savage:di}). In \cite{Brambilla:2005yk}, the 
obtained values are: $M_{\Xi^*_{cc}}-M_{\Xi_{cc}} = 120 \pm 40$ MeV 
and $M_{\Xi^*_{bb}}-M_{\Xi_{bb}} = 34 \pm 4$ MeV, which are 
consistent with the quenched lattice determinations of 
\cite{Flynn:2003vz,Lewis:2001iz,AliKhan:1999yb,Mathur:2002ce}.
Chiral corrections to the doubly  heavy baryon masses, strong decay widths and 
electromagnetic decay widths have been considered in \cite{Hu:2005gf}.

Also low-lying $QQQ$ baryons can be studied in a weak coupling framework.
Three quark states can combine in four color configurations: a singlet,
two octets and a decuplet, which lead to a rather rich dynamics
\cite{Brambilla:2005yk}. Masses of various $QQQ$ ground states have been 
calculated with a variational method in \cite{Jia:2006gw}: since baryons made of three 
heavy quarks have not been discovered so far, it may be important for future searches  
to remark that the baryon masses turn our to be lower 
than those generally obtained in strong coupling analyses.

\section{High-lying $Q\bar{Q}$}
High-lying $Q\bar{Q}$ states are expected to realize the hierarchy: $ m \gg  mv \sim \lQ
\gg mv^2$. Since we cannot measure directly $mv$, we do not know exactly   
where the border between low-lying and high-lying states lies.
This difficulty is reflected in the literature. A weak-coupling treatment for the lowest-lying 
bottomonium masses ($n=1$, $n=2$ states and the $\Upsilon(3S)$) works well  
at NNLO in \cite{Brambilla:2001fw} and at N$^3$LO in \cite{Penin:2005eu}.
The outcome is more ambiguous for the fine splittings of the 
bottomonium $1P$ levels in the NLO analysis of \cite{Brambilla:2004wu} and is positive only 
for the $\Upsilon(1S)$ mass in the N$^3$LO analysis of \cite{Beneke:2005hg}. 
In the weak-coupling regime, the magnetic-dipole hindered transition $\Upsilon(2S) \to \gamma\,\eta_b$ 
at leading order \cite{Brambilla:2005zw} does not agree with the experimental upper bound 
\cite{Artuso:2004fp}, while the ratios for different $n$ of the radiative decay widths 
$\Gamma(\Upsilon(nS) \to \gamma\,X)$ are better consistent with
the data if the $\Upsilon(1S)$ is assumed to be a weakly-coupled bound state 
and the $\Upsilon(2S)$ and $\Upsilon(3S)$ strongly coupled ones \cite{GarciaiTormo:2005bs}. 

Masses of high-lying quarkonia may be accessed by lattice calculations.
A recent unquenched QCD determination of the charmonium spectrum below the open flavor threshold 
with staggered sea quarks may be found in \cite{Gottlieb:2005me}. 
At present, bottomonium is too heavy to be implemented 
directly on the lattice. A solution is provided by NRQCD \cite{Lepage:1992tx}. 
Since the heavy-quark mass scale has been integrated out, 
for NRQCD on the lattice, it is sufficient to have a lattice spacing $a$ as coarse as $m \gg 1/a \gg mv$.
A price to pay is that, by construction, the continuum limit cannot be reached. 
Another one is that the NRQCD Lagrangian has to be supplemented by matching coefficients 
calculated in lattice perturbation theory, which encode the contributions from the 
heavy-mass energy modes that have been integrated out. A recent unquenched determination 
of the bottomonium spectrum with staggered sea quarks can be found in \cite{Gray:2005ur}.
The fact that all matching coefficients of NRQCD on the lattice are taken at their tree-level value 
induces a systematic error of order $\als v^2$ for the radial splittings 
and of order $\als$ for the fine and hyperfine splittings. 

At a scale $\mu$ such that $mv \sim \lQ \gg \mu \gg mv^2$, confinement 
sets in. Below threshold, the effective degrees of freedom are
$Q\bar Q$ states (in color singlet configuration) and light quarks. 
Without light quarks, the $Q\bar Q$ interaction would be simply described 
by a non-relativistic potential \cite{Brambilla:2000gk,Pineda:2000sz}.
The potential is in general a complex valued function admixture of perturbative 
terms, inherited from NRQCD, which encode high-energy contributions, 
and non-perturbative ones. The latter may be expressed in terms of Wilson loops 
and, therefore, are well suited for lattice calculations. 

The real part of the potential has been one of the first quantities to be calculated 
on the lattice (for a review see \cite{Bali:2000gf}).
In the last year, there has been some remarkable progress \cite{Schierholz:ICHEP06}.
In \cite{Koma:2006si}, the $1/m$ potential has been calculated for the first time.
The existence of this potential was first pointed out in the pNRQCD framework \cite{Brambilla:2000gk}. 
The lattice result shows that the potential has a  $1/r$ behaviour, which, in the charmonium case, 
is of the same size as the $1/r$ Coulomb tail of the static potential 
and, in the bottomonium one, is about 25\%. Therefore, if the $1/m$ potential is to be considered part 
of the leading-order quarkonium potential, then the latter would turn out to be, 
somewhat surprisingly, a flavor-dependent function.
In \cite{Koma:2006fw}, spin-dependent potentials have been calculated 
with unprecedented precision. In the long range, the spin-orbit potentials show, for the first 
time, deviations from the flux-tube picture of chromoelectric confinement \cite{Buchmuller:1981fr}, 
which is predicted in many models of the QCD vacuum \cite{Brambilla:1996aq}. Spin-spin and tensor potentials 
do not show sizeable long-range contributions. In the data,  this is reflected, for instance, 
by the smallness of the $P$-wave hyperfine splitting: the E835 experiment 
measures a $h_c$ mass $ M_{h_c} = 3525.8 \pm 0.2 \pm 0.2~{\rm MeV}$ ($\Gamma < 1~{\rm MeV}$) 
\cite{Patrignani:2004nf} and CLEO measures $M_{h_c} = 3524.4 \pm 0.6 \pm 0.4~{\rm MeV}$ \cite{Rubin:2005px}, 
both very close to the center-of-gravity mass $M_{\rm c.o.g.}(1P) = 3525.36 \pm 0.2 \pm 0.2~{\rm MeV}$.

The expectation value of the imaginary part of the potential provides the quarkonium width.
At the level of NRQCD, state-of-the art expressions for the decay widths may be 
found in \cite{Bodwin:2002hg,Ma:2002ev,Brambilla:2006ph}, at the level of pNRQCD in 
\cite{Brambilla:2001xy,Brambilla:2002nu,Brambilla:2003mu,Vairo:2003gh}.
Charmonium $P$-wave decay widths calculated in NRQCD \cite{Brambilla:2004wf}
and bottomonium $P$-wave decay widths calculated in pNRQCD \cite{Brambilla:2001xy,Vairo:2002nh}
are consistent with the most recent data \cite{Brambilla:2004wf}. 
In both cases, analyses have been performed at NLO in $\als$ and at leading order in the velocity expansion.

\section{Threshold states}
Most of the newly discovered resonances in the charmonium region are near or above 
threshold. A comprehensive review updated at January 2006 is \cite{Swanson:2006st}.
For states near or above threshold a general and systematic treatment does not exist so far. Most of the
existing analyses rely on models (e.g., for the coupling with two-mesons states, the Cornell coupled-channel model
\cite{Eichten:1978tg,Eichten:2004uh} and the $^3P_0$ model \cite{LeYaouanc:1972ae,Kalashnikova:2005ui}).
This makes it difficult to predict with precision even the masses of the states, 
which is usually not the case for states below threshold.

\subsection{Candidates for $c\bar{c}$ states} 
Among the recently discovered signals that may be ascribed to traditional $c\bar{c}$ 
bound states are the  $X(3940)$, $Z(3930)$ and $Y(3940)$.

The state labeled $X(3940)$ has been seen in  $e^+e^-\to J/\psi \, X$ by the BELLE collaboration \cite{Abe:2005hd}
with a mass $M_X = 3943 \pm 6 \pm 6~{\rm MeV}$ and a width $\Gamma < 52 ~{\rm MeV}$.
In \cite{Rosner:2005gf}, it has been suggested a $\eta_c(3S)$ interpretation for the state, 
but then the mass would be somewhat lower to what expected in potential models.
The state contributes to the large cross section of 
$e^+e^-\to J/\psi+ c\bar{c}$, which turns out to be $(82\pm 15\pm 14)\%$ of the 
total cross section $e^+e^-\to J/\psi+ X$ \cite{BELLE:EPS03}. Theoretical problems 
related to double charmonium production have been discussed in 
\cite{Likhoded:ICHEP06}. For recent progress see \cite{Zhang:2006ay}.

A state labeled $Z(3930)$ has been seen by the BELLE collaboration in 
$\gamma\gamma \to D\bar{D}$ \cite{Uehara:2005qd} with a mass 
$M_Z = 3929 \pm 5 \pm 2~{\rm MeV}$ and a width $\Gamma = 29\pm 10\pm 2 ~{\rm MeV}$.
So far, the properties of the state fit well with a $\chi_{c2}(2P)$ interpretation.

More enigmatic is the interpretation of the state $Y(3940)$
seen by the BELLE collaboration in $B \to K Y  \to K \omega J/\psi$ \cite{Abe:2004zs}.
The reported mass is $M_Y = 3943 \pm 11 \pm 13~{\rm MeV}$ and the width is 
$\Gamma = 87 \pm 22 \pm 26 ~{\rm MeV}$. The mass and width seem to fit with a radially excited 
$P$-wave charmonium, however the discovery decay process also suggests more exotic interpretations:
for instance, a tetraquark interpretation has been proposed in \cite{Maiani:2004vq}.

Finally, at the QWG meeting of this year \cite{Ye:QWG06}
and at this conference \cite{Lou:ICHEP06}, the BABAR collaboration 
reported a new signal seen in $e^+e^- \to \psi(2S) \pi\pi$.
If identified with a resonance, labeled $Y(4350)$, its mass would be 
$M_Y = 4354 \pm 16~{\rm MeV}$ and its decay width $\Gamma = 106 \pm 19~{\rm MeV}$.
So far, no theoretical interpretation has been put forward.

\subsection{Candidates for exotic states: $X(3872)$ and $Y(4260)$} 
Some of the newly discovered states allow for exotic interpretations. 
This is particularly the case for the $X(3872)$ and the $Y(4260)$.

The state $X(3872)$ has been discovered by BELLE in $B^\pm\to K^\pm X \to K^\pm \pi^+\pi^- J/\psi$ 
with $M_X = 3872.0 \pm 0.6 \pm 0.5~{\rm MeV}$ \cite{Choi:2003ue,Majumder:ICHEP06}, 
and confirmed by BABAR \cite{Aubert:2004ns,Lou:ICHEP06} that measures 
$M_X = 3871.3 \pm 0.6 \pm 0.1 ~{\rm MeV}$ in $B^-\to K^-\pi^+\pi^- J/\psi$ and 
$M_X = 3868.6 \pm 1.2 \pm 0.2 ~{\rm MeV}$ in $B^0\to K^0\pi^+\pi^- J/\psi$ \cite{Aubert:2005zh}.
The state has also been seen at the Tevatron 
in $p\bar{p}\to X \to \pi^+\pi^- J/\psi$ by CDF with a mass $M_X = 3871.3 \pm
0.7 \pm 0.4 ~{\rm MeV}$ \cite{Acosta:2003zx,Kreps:ICHEP06} 
and by D0 with a mass $M_X = 3871.8 \pm 3.1 \pm 3.0~{\rm MeV}$ \cite{Abazov:2004kp}. 
BELLE has an upper limit on the width: 
$\Gamma < 2.3$ MeV. The $X(3872)$ has been detected in four different decay modes:
$X \to$ $\pi^+\pi^-J/\psi$, which is the discovery mode, $X \to \pi^+\pi^-\pi^0 J/\psi$ \cite{Abe:2004sd}, 
$X \to \gamma J/\psi$ \cite{Abe:2005ix} and $X \to D^0 \bar{D}^0 \pi^0$ \cite{Gokhroo:2006bt}.
The last one is likely to be the dominant decay mode: in \cite{Aubert:2005vi}, 
it was found that ${\cal B}(X \to \pi^+\pi^- J/\psi) > 4.2\%$ at 90\% C.L., which,  
combined with the ratio ${\cal B}(X \to D^0 \bar{D}^0 \pi^0)/{\cal B}(X\to
  \pi^+\pi^-J/\psi) = 9.4^{+3.6}_{-4.3}$ measured in \cite{Gokhroo:2006bt}, gives 
${\cal B}(X \to D^0 \bar{D}^0 \pi^0) > 40^{+15}_{-20}\%$. One should notice that 
BELLE finds a threshold enhancement peak in the $D^0 \bar{D}^0 \pi^0$ invariant mass 
at $3875.4 \pm 0.7^{+1.2}_{-2.0}$ MeV, which is 2.0 $\sigma$ larger than the world-average 
mass of the $X(3872)$. The decay mode $X\to \gamma J/\psi$ implies that the 
$X(3872)$ has positive charge conjugation. Analyses of angular distributions 
performed by BELLE \cite{Abe:2005iy} and CDF \cite{Kravchenko:2006qx}  favor 
a spin parity assignment $1^+$. The ratio ${\cal B}(X \to \pi^+\pi^-\pi^0 J/\psi)/{\cal B}(X\to
  \pi^+\pi^-J/\psi) = 1.0 \pm 0.4 \pm 0.3$ measured by BELLE \cite{Abe:2005ix}
suggests that the  $X(3872)$ is a mixture of isospin $I=1$ and $I=0$ states.
The substantial $I=1$ component requires that the $X(3872)$ contains 
$u\bar{u}$/$d\bar{d}$ pairs in addition to hidden charm, which thus 
qualifies it as a four-quark state \cite{Voloshin:2006wf}.
Hence, most recently, the majority of theoretical studies has  
analyzed the $X(3872)$ as a four-quark state with $J^{PC}$ quantum numbers $1^{++}$.

Three equivalent quark-pair configurations are possible for a four-quark state of the type 
$c\bar{c}q\bar{q}$ ($q$ stands for a generic light quark).
They have been all exploited in the literature. However, the resulting models are not equivalent, 
because different dynamics are attributed to different configurations.
In \cite{Hogaasen:2005jv}, it is assumed that $X \sim (c\bar{c})^8_{S=1}   (q\bar{q})^8_{S=1}$, 
i.e. that the dominant Fock-space component contains a $c\bar{c}$ pair and a $q\bar{q}$ pair
in a color octet configuration with spin 1. This configuration is equivalent 
to $ X \sim (c\bar{q})^1_{S=0}   (q\bar{c})^1_{S=1}
+ (c\bar{q})^1_{S=1}   (q\bar{c})^1_{S=0}$, which is the molecular picture that we will discuss later.
Calculations are based on the phenomenological interaction Hamiltonian 
$H = - \sum_{ij} C_{ij} \; T^a \otimes T^a \; \bfsigma \otimes \bfsigma$.
It is expected that decays into charmonium with light pseudoscalar mesons are suppressed with respect 
to decays into charmonium with light vector mesons; moreover, 
two neutral states made of $cu\bar{c}\bar{u}$ and $cd\bar{c}\bar{d}$
and two charged ones made of $cu\bar{c}\bar{d}$ and $cd\bar{c}\bar{u}$ should exist.

In \cite{Maiani:2004vq}, it is assumed that $X \sim (cq)^{\bar{3}}_{S=1}   (\bar{c}\bar{q})^{3}_{S=0}
+ (cq)^{\bar{3}}_{S=0}   (\bar{c}\bar{q})^{3}_{S=1}$. 
Here, the clustering of quark pairs in tightly bound color triplet diquarks 
is not induced by a scale separation, like in the doubly heavy baryon case discussed above, 
but is a dynamical assumption of the model. Predictions are based on the phenomenological interaction Hamiltonian 
$ H = \sum_{ij} \kappa_{ij} \; \bfsigma \otimes \bfsigma$. 
In particular, the model predicts the existence of two neutral states made of $cu\bar{c}\bar{u}$ and $cd\bar{c}\bar{d}$
and of two charged ones. The mass difference between the two neutral states should be 
$\Delta M_X = 2 (m_d-m_u)/ \cos(2\theta) \approx (8\pm 3)$ MeV
if the mixing angle $\theta$ is fixed on $\Gamma(X \to \pi^+\pi^-\pi^0 J/\psi)/\Gamma(X\to\pi^+\pi^-J/\psi)$.
Since $B^\pm \to K^\pm X$ and $B^0 \to K^0 X$ produce $cu\bar{c}\bar{u}$
and $cd\bar{c}\bar{d}$ in different amount, the mass difference should be
seen in the two processes. The BABAR result \cite{Aubert:2005zh}, 
$\Delta M_X = (2.7 \pm 1.3 \pm 0.2)$ MeV, is not conclusive.
Searches for charged partners of the $X(3872)$ have been negative so far, 
while the $2^{++}$ partner may be consistent with the $Y(3940)$.
In the same framework, a tetraquark interpretation of the $D_s$ particles 
has been also advanced.

In \cite{Tornqvist:1993ng,Swanson:2003tb}, it is assumed that
$X \sim (c\bar{q})^1_{S=0}   (q\bar{c})^1_{S=1} 
+ (c\bar{q})^1_{S=1}   (q\bar{c})^1_{S=0}$ $ \sim D^0\,\bar{D}^{*\,0} + D^{*\,0}\, \bar{D}^0$, 
i.e. that the dominant Fock-space component of the $X(3872)$ is a $D^0\, \bar{D}^{*\,0}$ and 
$D^{*\,0} \, {\bar D}^0$ molecule; small short-range components 
of the type $(c\bar{c})^1_{S=1}   (q\bar{q})^1_{S=1}$ $\sim$ $J/\psi \,\rho, \omega$ are
included as well. Predictions depend on the adopted phenomenological
Hamiltonian, which typically contains, in the short range ($\sim 1/\lQ$), 
potential-type interactions among the quarks and,  
in the long range ($\sim 1/m_\pi$), the one-pion exchange. 
The prediction $\Gamma(X\to \pi^+\pi^-J/\psi) \approx \Gamma(X\to
  \pi^+\pi^-\pi^0 J/\psi)$ made in \cite{Swanson:2003tb} turned out to be
  consistent with the BELLE result \cite{Abe:2005ix}. 
However, another prediction, $\Gamma(X\to \pi^+\pi^-J/\psi) \approx 20 \,\Gamma(X\to
  D^0 \bar{D}^0 \pi^0)$, is two orders of magnitude far from 
the data  \cite{Gokhroo:2006bt}. Not necessarily this points to a failure of
the molecular model, but possibly to a smaller $J/\psi \,\rho$ component in the $X(3872)$ Fock space.

In \cite{Pakvasa:2003ea,Voloshin:2003nt,Braaten:2003he}, 
it is assumed not only that the $X(3872)$ is a $D^0 \, \bar{D}^{*\,0}$ and  
${\bar D}^0 \, D^{*\,0}$ molecule, but also that it is loosely bound, 
i.e. that the following hierarchy of scales is realized:
$\lQ \gg m_\pi$ $\gg m_\pi^2/M_{D_0}$ 
$\approx$  $10 ~{\rm MeV}$ $\gg E_{\rm binding}$. Indeed, the binding energy, 
$E_{\rm binding}$, which may be estimated from $M_X - (M_{D^{*\,0}}+M_{D^{0}})$,
is very close to zero, i.e. much smaller than the natural scale 
$m_\pi^2/M_{D_0}$. This is also the case when using the new CLEO determination 
of the $D_0$ mass, $M_{D_0} = 1864.85\pm 0.15\pm 0.20$ MeV, \cite{Skwarnicki:ICHEP06}. The main uncertainty comes from the 
$X(3872)$ mass. Systems with a short-range interaction 
and a large scattering length have universal properties that may be exploited: 
in particular, production and decay amplitudes factorize in a short-range 
and a long-range part, where the latter depends only on one single parameter, 
the scattering length.  A universal property that fits well 
with the observed large branching fraction of the $X(3872)$ decaying into 
$D^0 \bar{D}^0 \pi^0$ is 
${\cal B}(X\to D^0 \bar{D^{0}}\pi^0) \approx {\cal B}(D^{*\,0} \to D^{0} \pi^0) \approx 60\%$.
This supports the view that the difficulties met in \cite{Swanson:2003tb} 
to account for this decay channel may be specific of that model and not 
of the molecular picture in general.

The state $Y(4260)$ has been discovered by BABAR in the radiative return process 
$e^+e^- \to \gamma \pi^+\pi^- J/\psi$ with mass $M_Y = 4259 \pm 8^{+2}_{-6}~{\rm MeV}$
and width $\Gamma =  88 \pm 23 ^{+6}_{-4}~{\rm MeV}$ 
\cite{Aubert:2005rm,Lou:ICHEP06}, and seen in the same process by BELLE
with mass $M_Y = 4295 \pm 10^{+11}_{-5}~{\rm MeV}$ and width 
$\Gamma =  133 \pm 26 ^{+13}_{-6}~{\rm MeV}$ \cite{Olsen:QWG06,Majumder:ICHEP06}
and by CLEO with mass $M_Y = 4284^{+17}_{-16}\pm 4~{\rm MeV}$ and width $\Gamma =  73^{+39}_{-25} \pm 5~{\rm MeV}$
\cite{He:2006kg}. CLEO has also confirmed the existence of an enhancement in the 
$\pi^+\pi^-J/\psi$ cross section at 4260 MeV in a measurement of direct $e^+e^-$ annihilation 
at $\sqrt{s}=4040$, $4160$ and $4260$ MeV \cite{Coan:2006rv,Shipsey:ICHEP06}. 
The $Y(4260)$ $J^{PC}$ quantum numbers are $1^{--}$.
The $\pi^+\pi^-$ spectrum  does not show signs of $f_0(980)$ or $f_0(600)$ 
\cite{Coan:2006rv}. BABAR measures ${\cal B}(Y \to D \bar{D})/{\cal B}(Y \to 
    \pi^+\pi^- J/\psi) < 7.6$  ($\approx 500$ for $\psi(3770)$, which suggests an exotic interpretation 
for the $Y(4260)$) \cite{Aubert:2006mi,Lou:ICHEP06}; BELLE sees a strong drop and local minimum of the
$D^{*\,+} D^{*\,-}$ invariant mass at 4260 MeV in $e^+e^- \to \gamma D^{*\,+} D^{*\,-}$  
\cite{BELLE:2006fj,Majumder:ICHEP06}.

Many interpretations have been suggested for the $Y(4260)$:
$Y \sim \psi(4S)$ \cite{Llanes-Estrada:2005hz}, $Y \sim \Lambda_c \bar{\Lambda}_c$ \cite{Qiao:2005av},
$Y \sim [(cs)^{\bar{3}}_{S=0}   (\bar{c}\bar{s})^{3}_{S=0}]_{\rm P-wave}$ \cite{Maiani:2005pe}, 
$Y \sim [(cq)^{\bar{3}}_{S=0}   (\bar{c}\bar{q})^{3}_{S=0}]_{\rm P-wave}$ \cite{Zhu:2005hp,Ebert:2005nc} 
$Y \sim \chi_{c1}\rho$  \cite{Liu:2005ay}, $Y\sim \chi_{c1}\omega$  \cite{Yuan:2005dr} and 
$Y$ as a charmonium hybrid \cite{Zhu:2005hp,Kou:2005gt,Close:2005iz}.
In particular, if interpreted as a charmonium hybrid, one may rely on the heavy-quark expansion 
and on lattice calculations to study its properties. Decays into $D^{(*)} \bar{D}^{(*)}$ should be suppressed, since 
they are forbidden at leading order in the heavy-quark expansion
\cite{Kou:2005gt} (in the tetraquark picture  \cite{Maiani:2005pe} the most natural decay should be into $D_s\bar{D_s}$).
This is in agreement with the upper limit on $Y \to D \bar{D}$ reported by BABAR and calls for 
an understanding of the signal seen by BELLE in $e^+e^- \to \gamma D^{*\,+} D^{*\,-}$.
It is suggestive also that, according to lattice calculations \cite{Juge:2002br},  
the lowest hybrid state is expected to contain a pseudoscalar 
color-octet quark-antiquark pair and gluons, whose quantum numbers are those of the electric cloud in a 
diatomic $\Pi_u$ molecule, so that, indeed, the system has $J^{PC}$ numbers $1^{--}$, like the $Y$.

\subsection{X and Y in the bottomonium region} 
States analogous to the $X(3872)$ and the $Y(4260)$ could also exist in the
bottomonium region. Their finding could provide a confirmation of their  
interpretations in a more controlled framework for the theory.
Possibilities for searches of these states have been discussed in \cite{Hou:2006it}.

In the framework of the molecular picture, in \cite{AlFiky:2005jd},   
a  $1^{++}$ $B^0\, \bar{B}^{*\,0}$ and $B^{*\,0} \, {\bar B}^0$ molecular state is predicted at   
$10604$ MeV. In \cite{Ebert:2005nc}, the  $1^{++}$ 
$(bq)^{\bar{3}}_{S=1}   (\bar{b}\bar{q})^{3}_{S=0}
+ (bq)^{\bar{3}}_{S=0}   (\bar{b}\bar{q})^{3}_{S=1}$ tetraquark  is predicted at  $10492$ MeV.
The lattice determination \cite{Manke:2001ft} predicts the lowest bottomonium hybrid at a mass of $11020 \pm 180$ MeV
(solving the Schr\"odinger equation for the $\Pi_u$ static energy calculated
in \cite{Juge:2002br} at intermediate distances gives a mass of about 10750 MeV).
In \cite{Ebert:2005nc}, the  $1^{--}$ $[(bq)^{\bar{3}}_{S=0}   (\bar{b}\bar{q})^{3}_{S=0}]_{\rm P-wave}$ 
tetraquark is predicted at 10807 MeV  and the $1^{--}$ $[(bs)^{\bar{3}}_{S=0} 
  (\bar{b}\bar{s})^{3}_{S=0}]_{\rm P-wave}$ one at 11002 MeV.

\section{Conclusions}
A new era of heavy hadron spectroscopy has begun.
It has been initiated experimentally by the B-factories, CLEO, BES, 
and the Tevatron experiments. They have provided measurements with 
unprecedented precision and shown evidence of new, perhaps exotic, states, and 
new production and decay mechanisms. It will continue with the BES upgrade, 
the LHC and GSI experiments and possibly with a new tau-charm factory and a super B factory. 
The experimental renaissance has been accompanied by an analogous one  
in the theory of heavy hadrons, whose language is rapidly 
becoming that one of effective field theories and lattice gauge theories.
For many observables systematic and controlled expansions exist  
that lead to definite theoretical results. The construction of similar expansions 
for states near or above threshold still remains a challenge.

\section*{Acknowledgements}
I acknowledge the financial support obtained inside the Italian MIUR program 
``incentivazione alla mobilit\`a di studiosi stranieri e italiani residenti all'estero''.

\end{document}